\documentclass{emulateapj}

\shorttitle{The bright end of the $z \sim$ 5 QLF based on SDSS-WISE selection}
\shortauthors{Yang et al.}

\begin{document}


\title{A survey of luminous high-redshift quasars with SDSS and WISE II. the bright end of the quasar luminosity function at \lowercase{$z \sim$ 5}}


\author{Jinyi Yang\altaffilmark{1,2}, Feige Wang\altaffilmark{1,2}, Xue-Bing Wu\altaffilmark{1,3}, Xiaohui Fan\altaffilmark{2,3}, Ian D. McGreer\altaffilmark{2}, Fuyan Bian\altaffilmark{4,5}, Weimin Yi\altaffilmark{6,7}, Qian Yang\altaffilmark{1,2}, Yanli Ai\altaffilmark{8}, Xiaoyi Dong\altaffilmark{1}, Wenwen Zuo\altaffilmark{9}, Richard Green\altaffilmark{2}, Linhua Jiang\altaffilmark{3}, Shu Wang\altaffilmark{1}, Ran Wang\altaffilmark{3}, Minghao Yue\altaffilmark{1}}

\altaffiltext{1}{Department of Astronomy, School of Physics, Peking University, Beijing 100871, China}
\altaffiltext{2}{Steward Observatory, University of Arizona, 933 North Cherry Avenue, Tucson, AZ 85721, USA}
\altaffiltext{3}{Kavli Institute for Astronomy and Astrophysics, Peking University, Beijing 100871, China}
\altaffiltext{4}{Research School of Astronomy and Astrophysics, Australian National University, Weston Creek, ACT 2611, Australia}
\altaffiltext{5}{Stromlo Fellow}
\altaffiltext{6}{Yunnan Observatories, Chinese Academy of Sciences, Kunming 650011,China}
\altaffiltext{7}{Key Laboratory for the Structure and Evolution of Celestial Objects, Chinese Academy of Sciences, Kunming 650011,China}
\altaffiltext{8}{School of Astronomy and Space Science, Sun Yat-Sen University, Guangzhou 510275, China}
\altaffiltext{9}{Shanghai Astronomical Observatory, Chinese Academy of Sciences, Shanghai 200030, China}



\begin{abstract}
This is the second paper in a series on a new luminous $z \sim 5$ quasar survey using optical and near-infrared colors. Here we present a new determination of the bright end of the quasar luminosity function (QLF) at $z \sim$ 5. Combined our 45 new quasars with previously known quasars that satisfy our selections, we construct the largest uniform luminous $z \sim$ 5 quasar sample to date, with 99 quasars in the range 4.7 $\le z <$ 5.4 and $-29 < M_{1450} \le -26.8$, within the Sloan Digital Sky Survey (SDSS) footprint. We use a modified 1/V$_{\rm a}$ method including flux limit correction to derive a binned QLF, and we model the parametric QLF using maximum likelihood estimation. With the faint-end slope of the QLF fixed as $\alpha = -2.03$ from previous deeper samples, the best fit of our QLF gives a flatter bright end slope $\beta = -3.58\pm0.24$ and a fainter break magnitude $M^{*}_{1450} = -26.98\pm0.23$ than previous studies at similar redshift. Combined with previous work at lower and higher redshifts, our result is consistent with a luminosity evolution and density evolution (LEDE) model. Using the best fit QLF, the contribution of quasars to the ionizing background at $z \sim$ 5 is found to be 18\% $-$ 45\% with a clumping factor $C$ of 2 $-$ 5. Our sample suggests an evolution of radio loud fraction with optical luminosity but no obvious evolution with redshift.

\end{abstract}





\section{Introduction}
 Quasars comprise the most luminous class of non-transient objects in the universe. Characterizing their population and evolution is the critical tool to constrain directly the formation and evolution of supermassive black holes (SMBHs) across cosmic time. The fundamental way to characterize these objects is through the evolution of their number density with luminosity and redshift,  namely the quasar luminosity function (QLF). The QLF and its cosmological evolution have been a key focus of quasar studies for half a century. \cite{schmidt68} first determined the evolution of the quasar population and found the first evidence for a significant increase of the quasar number density with redshift in both radio and optical bands. More recently, based on measurements of the QLF from several successful surveys, such as the 2dF Quasar Redshift Survey \citep{boyle00, croom04}, COMBO-17 \citep{wolf03}, the 2dF-SDSS LRG and QSO survey \cite[2SLAQ;][]{richards05}, the SDSS Faint Quasar Survey \citep{jiang06}, the VIMOS-VLT Deep Survey \cite[VVDS;][]{bongiorno07}, SDSS and 2SLAQ \citep{croom09} and BOSS DR9 \citep{ross13}, the QLF, especially in optical bands, has been well characterized at low to intermediate redshifts. The QLF can be parameterized with a double power law shape and pure luminosity evolution for quasars at redshifts up to $z$  = 2 \citep{boyle00, croom04}. The bright end slope at low redshift, the effect of $''$cosmic downsizing$''$ and the density peak of quasars at 2 $< z <$ 3 \citep{richards06, brown06, jiang06, croom09} have been confirmed by many subsequent investigations. The measurements based on large samples from BOSS yield a QLF evolution best fit by a luminosity evolution and density evolution (LEDE) model at 2 $< z <$ 3.5 \citep{ross13}. In their work, the bright end slope does not evolve with redshift and is different from the result of \cite{richards06}, which suggested a flatter bright end slope at high redshift than that at low redshift. To better determine the evolution of QLF parameters, a wider redshift range is needed.

Towards higher redshift, quasars are important tracers of the structure and evolution of the early Universe, the evolution of the intergalactic medium (IGM), the growth of SMBHs and co-evolution of SMBHs and host galaxies at early epochs. Observations of the Gunn-Peterson effect using absorption spectra of quasars at $z \gtrsim$ 5.7 have established $z \sim$ 6 as the end of cosmic reionization, when the IGM is rapidly transforming from largely neutral to completely ionized \citep{fan06}. \cite{becker15} find evidence for UV background fluctuations at $z\sim$ 5.7 in excess of predictions from a single mean-free-path model, which indicates that reionization is not fully complete at that redshift.  \cite{mcgreer15} suggest that reionization is just completing at $z\sim$ 6, possibly with a tail to $z\sim$ 5.5. Therefore, in the post-reionization epoch, the QLF at $z \gtrsim 5$ is needed to estimate the contribution of quasars to the ionizing background during and after the reionization epoch. Although quasars are not likely to be the dominant source of ionizing photons \citep{fan01a, willott10b, mcgreer13}, their exact contribution is still highly uncertain. In addition, $z\sim 5$ quasar absorption spectra can be used to constrain the physical conditions of the IGM in this key redshift range, and provide the basic boundary conditions for models of reionization, such as the evolution of IGM temperature, photon mean free path, metallicity and the impact of helium reionization \citep{bolton12}.

However, high redshift quasars are very rare, especially at $z >$ 5. Although more than 300,000 quasars are known, only $\sim$200 of them are at $z >$ 5. Therefore, QLF measurements at high redshift  still have large uncertainties.
From the combination of SDSS DR7 quasars and the Stripe 82 (S82) faint quasar sample, \cite{mcgreer13}  provided the most complete measurement of the $z \sim$5 QLF so far, especially at the faint end. A factor of 2 greater decrease in the number density of luminous quasars from $z$  = 5 to 6 than that from $z$  = 4 to 5 was claimed \citep[][hereafter M13]{mcgreer13}.
However, their work focused on the faint end; there are only 8 quasars with $M_{1450} < -27.3$ in the sample. 

A survey described in this series of papers aims at finding more luminous quasars at 4.7 $< z <$ 5.5,  which allows a better determination of the  bright end QLF and a better constraint on the quasar evolution model at high redshift. \citet[][hereafter Paper I]{wang16} presented a new selection using SDSS and the Wide-field Infrared Survey Explorer (WISE) optical/NIR colors. In this followup paper, we report our measurement of the bright end $z \sim$ 5 QLF using the quasar sample selected by the method presented in Paper I. The outline of our paper is as follows. In Section 2, we briefly review the quasar candidate selection and the spectroscopic observations of these candidates. The survey completeness will be presented in Section 3. In this section, we use a quasar color model (M13) to quantify our selection completeness and to correct the incompleteness due to the ALLWISE detection flux limit and spectral coverage. We then calculate the binned luminosity function and fit our data using a maximum likelihood estimator in Section 4. We also study the evolution of the QLF and compare our results with previous work in this section. In Section 5, we discuss the contribution of $z \sim$ 5 quasars to the ionizing background and the radio loud fraction of our quasar sample. We summarize our main results in Section 6. In this paper, we adopt a $\Lambda$CDM cosmology with parameters $\Omega_{\Lambda}$ = 0.728, $\Omega_{m}$ = 0.272, $\Omega_{b}$ = 0.0456, and H$_{0}$ = 70 $km s^{-1} Mpc^{-1}$ \citep{komatsu09} for direct comparison with the result in M13. Photometric data from the SDSS are in the SDSS photometric system \citep{Lupton99}, which is almost identical to the AB system at bright magnitudes; photometric data from ALLWISE are in the Vega system. All SDSS data shown in this paper are corrected for Galactic extinction.
 
\section{A Large Sample of Luminous Quasars\\ at $z \sim$ 5}
\subsection{Quasar selection and Spectroscopic observations}

Our SDSS+WISE selection technique and spectroscopic follow-up observations were discussed in detail in Paper I. Here we briefly review the basic steps. At $z \sim$ 5, most quasars are undetectable in $u$-band and $g$-band because of the presence of strong Lyman limit systems (LLSs), which are optically thick to the UV continuum radiation from quasars \citep{fan99b}. The Ly$\alpha$ absorption systems also begin to dominate in the $r$-band and Ly$\alpha$ emission moves to the $i$-band. Therefore, the $r-i$/$i-z$ color-color diagram was often used to select $z \sim$ 5 quasar candidates in previous studies \citep[][M13]{fan99b, richards02}. However, with increasing redshift,  the $i-z$ color also becomes increasingly red and most $z >$ 5.1 quasars enter the M star locus in the $r-i$/$i-z$ color-color diagram, which makes it difficult to find $z >$ 5.1 quasars with only the optical colors. Therefore, we added near-infrared colors from WISE photometry data in our selection. We used typical $u$, $g$ drop-out methods but more relaxed $r-i$/$i-z$ cuts to select candidates from the SDSS DR10 database. Then we cross-matched our candidates with the ALLWISE database using a $2''$ match radius and used $z$-W1/W1-W2 cuts to remove more star contaminations by the following criteria. The exact selection criteria are given in Paper I.
\begin{equation}
z-W1 > 2.5
\end{equation}
\begin{equation}
W1-W2 > 0.5
\end{equation}
\begin{equation}
W1 < 17.0, \sigma_{W2} < 0.2
\end{equation}
\begin{equation}
z-W1 > 2.8~ or~ W1-W2 > 0.7, if~ i-z>0.4
\end{equation}
 
We constructed our main luminous quasar candidate sample by limiting the SDSS $z$ band magnitudes to brighter than 19.5, and selected a total of 420 luminous $z \sim$ 5 quasar candidates. We removed 78 known quasars, one known dwarf and 231 candidates with suspicious detections, such as multiple peaked objects or being affected by bright star artifacts. We visually inspected images of each candidate and removed those 231 candidates. We selected 110 candidates with high image quality as our main candidate sample. Our spectroscopic follow-up campaign started in October 2013. We observed 99 candidates from our main sample with the Lijiang 2.4m telescope (LJT) and Xinglong 2.16m telescope in China, the Kitt Peak 2.3m Bok telescope and 6.5m MMT telescope in the U.S., as well as the 2.3m ANU telescope in Australia. 64 (64.6\%) candidates have been identified as high-redshift quasars in the redshift range $4.4 \lesssim z \lesssim 5.5$. As discussed in Paper I, due to the serious contamination from M-type stars, there is a gap in the previously published quasar redshift distribution at 5.2 $< z <$ 5.7 with only 33 published quasars ins this redshift range. Among our 64 newly identified quasars from main candidates sample, 9 quasars are at 5.2 $< z <$ 5.7, which represents an increase of 27\% in the number of known quasars in this redshift range. The details of spectroscopic observation and data reduction are also given in Paper I. 

\subsection{Quasar sample}

The redshifts of newly identified quasars are measured from $\rm Ly \alpha$, N\,{\sc v}, O\,{\sc i}/Si\,{\sc ii}, C\,{\sc ii}, Si\,{\sc iv} and C\,{\sc iv} emission lines (any available) by an eye-recognition assistant for quasar spectra software \cite[ASERA;][]{yuan13}. The typical redshift error is about 0.05 for Ly$\alpha$-based redshift measurement and will be less for that based on more emission lines. We calculate $M_{1450}$ in the AB system by fitting a power-law continuum $f_\nu \sim  \nu^{\alpha_{\nu}}$ to the spectrum for each quasar. We assume an average quasar UV continuum slope of $\alpha_{\nu}=-0.5$ \citep{vandenberk01} (See details in Paper I). Our 64 new quasars from the luminous quasar candidate sample are within the absolute magnitude range $-29 < M_{1450} < -26.4$. We calculate $M_{1450}$ for previous known quasars using the same method. The known quasars are from the SDSS DR7 and DR12 quasar catalogs \citep[][P{\~a}ris et al. in prep]{schneider10} , \cite{mcgreer13} and \cite{schneider91}. When we removed the known quasars from our quasar candidate sample, we missed two known quasars. These two quasars from \cite{mcgreer09} and \cite{schneider91} were also spectroscopically observed by us, and thus we use our new spectra to do the $M_{1450}$ calculation.

For the QLF determination, we define our sample of $z\sim$ 5 luminous quasars as follows:
\begin{itemize}
\item Quasars in the redshift range 4.7 $\le  z <$ 5.4. Our selection criteria yield low completeness at redshifts lower than 4.7 or higher than 5.4. The former is caused by the drop of W1-W2 color ( See Fig. 3 in Paper I), and the latter is caused by our $r-i/i-z$ limit. Therefore, we restrict our sample to the range 4.7 $\le  z <$ 5.4 (See details in Section 3.2).
\item Quasars in the luminosity range $M_{1450} \le -26.8$. Our selection criteria yield a low completeness in the region with $z >$ 5 and $-26.8 < M_{1450} < -26.4$. The mean completeness in this region is $\sim$ 4\%. That is caused by our SDSS magnitude limit of  $z < 19.5$. Therefore, we limit our sample to $M_{1450} \le -26.8$.
\item Our selection covers the whole SDSS footprint without masked regions, which is a 14555 square degree field.
\end{itemize}

Based on the criteria above, there are 45 newly identified quasars in the sample of Paper I, and another 54 previously known quasars that satisfy our selection criteria. This is the final complete $z\sim 5$ luminous quasar sample that we will use to determine the $z\sim 5$ QLF. Figure 1 shows the redshift and $M_{1450}$ distributions of both our newly identified luminous quasars and known quasars. Three of our new quasars are more luminous than any previously known quasars at $z >$ 5.  It is obvious that our discovery significantly expands the $z \sim$ 5 luminous quasar sample. Table 1 lists all 99 quasars in our sample used for the QLF determination. 

\LongTables
\begin{deluxetable*}{llllllclc}
 \centering
 \tablecaption{$z\sim5$ quasar sample used for QLF measurement.}
 \tablewidth{0pt}
 \tablehead
{
  \colhead{name} &
  \colhead{$r_{\rm SDSS}$} &
  \colhead{$i_{\rm SDSS}$} &
  \colhead{$z_{\rm SDSS}$} &
  \colhead{W1} &
  \colhead{W2} &
  \colhead{redshift} &
  \colhead{$M_{1450}$} &
  \colhead{Notes\tablenotemark{a}}
 }
   \startdata
J000851.43+361613.49 & 21.45$\pm$0.08 & 19.50$\pm$0.02 & 19.20$\pm$0.05 & 16.05$\pm$0.05 & 15.37$\pm$0.09 & 5.17  & $-$27.41  & Paper I \\
J001115.24+144601.80 & 19.48$\pm$0.02 & 18.17$\pm$0.02 & 18.03$\pm$0.03 & 15.29$\pm$0.04 & 14.69$\pm$0.06 & 4.96  & $-$28.43  & DR12 \\
J002526.84$-$014532.51 & 19.58$\pm$0.02 & 18.03$\pm$0.02 & 17.85$\pm$0.02 & 14.80$\pm$0.03 & 14.16$\pm$0.05 & 5.07  & $-$28.70  & Paper I \\
J005527.19+122840.67 & 20.23$\pm$0.03 & 18.71$\pm$0.02 & 18.66$\pm$0.04 & 15.45$\pm$0.05 & 14.95$\pm$0.09 & 4.70  & $-$27.52  & Paper I \\
J011614.30+053817.70 & 21.57$\pm$0.09 & 19.87$\pm$0.03 & 19.22$\pm$0.06 & 16.37$\pm$0.07 & 15.76$\pm$0.13 & 5.33  & $-$27.73  & Paper I \\
J012247.35+121624.06 & 22.25$\pm$0.14 & 19.37$\pm$0.03 & 19.27$\pm$0.06 & 15.59$\pm$0.05 & 14.91$\pm$0.07 & 4.79  & $-$26.86  & Paper I \\
J013127.34$-$032100.19 & 20.15$\pm$0.04 & 18.46$\pm$0.02 & 18.01$\pm$0.03 & 14.58$\pm$0.03 & 13.84$\pm$0.04 & 5.18  & $-$28.44  & Paper I \\
J014741.53$-$030247.88 & 20.08$\pm$0.03 & 18.53$\pm$0.02 & 18.21$\pm$0.02 & 14.86$\pm$0.03 & 14.32$\pm$0.05 & 4.75  & $-$27.84  & Paper I \\
J015533.28+041506.74 & 21.70$\pm$0.10 & 19.97$\pm$0.03 & 19.26$\pm$0.06 & 16.33$\pm$0.07 & 15.19$\pm$0.10 & 5.37  & $-$27.10  & Paper I \\
J015618.99$-$044139.88 & 20.77$\pm$0.04 & 19.10$\pm$0.02 & 19.13$\pm$0.05 & 15.36$\pm$0.04 & 14.69$\pm$0.06 & 4.94  & $-$27.24  & Paper I \\
J021624.16+230409.47 & 21.26$\pm$0.06 & 19.78$\pm$0.03 & 19.32$\pm$0.06 & 16.56$\pm$0.08 & 15.73$\pm$0.15 & 5.26  & $-$27.25  & Paper I \\
J021736.76+470826.48 & 20.55$\pm$0.05 & 18.96$\pm$0.02 & 18.88$\pm$0.05 & 15.76$\pm$0.05 & 15.14$\pm$0.08 & 4.81  & $-$27.10  & Paper I \\
J022055.59+473319.34 & 20.07$\pm$0.03 & 18.34$\pm$0.01 & 18.31$\pm$0.03 & 15.19$\pm$0.04 & 14.62$\pm$0.06 & 4.82  & $-$27.85  & Paper I \\
J024601.95+035054.12 & 21.05$\pm$0.05 & 19.28$\pm$0.02 & 19.36$\pm$0.05 & 16.67$\pm$0.07 & 15.74$\pm$0.14 & 4.96  & $-$27.00  & Paper I \\
J025121.33+033317.42 & 20.80$\pm$0.04 & 19.04$\pm$0.03 & 19.06$\pm$0.05 & 15.64$\pm$0.04 & 14.93$\pm$0.07 & 5.00  & $-$26.89  & Paper I \\
J030642.51+185315.85 & 19.89$\pm$0.03 & 17.96$\pm$0.01 & 17.47$\pm$0.02 & 14.31$\pm$0.03 & 13.46$\pm$0.04 & 5.36  & $-$28.99  & Paper I \\
J032407.69+042613.29 & 20.39$\pm$0.04 & 19.03$\pm$0.02 & 19.15$\pm$0.06 & 15.72$\pm$0.05 & 15.13$\pm$0.09 & 4.72  & $-$27.19  & Paper I \\
J045427.96$-$050049.38 & 19.91$\pm$0.03 & 18.59$\pm$0.03 & 18.39$\pm$0.03 & 15.09$\pm$0.03 & 14.53$\pm$0.05 & 4.93  & $-$27.61  & Paper I \\
J065330.25+152604.71 & 21.27$\pm$0.06 & 19.48$\pm$0.02 & 19.39$\pm$0.07 & 16.65$\pm$0.11 & 15.79$\pm$0.16 & 4.90  & $-$27.09  & Paper I \\
J073103.13+445949.43 & 20.66$\pm$0.04 & 19.06$\pm$0.02 & 19.07$\pm$0.05 & 15.82$\pm$0.05 & 15.30$\pm$0.09 & 4.98  & $-$27.29  & DR12 \\
J073231.28+325618.33 & 20.26$\pm$0.03 & 18.82$\pm$0.01 & 18.62$\pm$0.03 & 15.46$\pm$0.04 & 14.92$\pm$0.08 & 4.76  & $-$27.60  & Paper I \\
J074154.72+252029.65\tablenotemark{b} & 20.49$\pm$0.03 & 18.45$\pm$0.02 & 18.36$\pm$0.02 & 14.78$\pm$0.03 & 13.81$\pm$0.04 & 5.21  & $-$28.31  & McGreer2009 \\
J074749.18+115352.46 & 20.44$\pm$0.03 & 18.67$\pm$0.02 & 18.27$\pm$0.03 & 14.64$\pm$0.03 & 13.79$\pm$0.04 & 5.26  & $-$28.04  & Paper I \\
J075907.58+180054.71 & 20.95$\pm$0.04 & 19.12$\pm$0.02 & 19.11$\pm$0.04 & 16.07$\pm$0.07 & 15.39$\pm$0.11 & 4.78  & $-$27.05  & DR12 \\
J080306.19+403958.96 & 20.58$\pm$0.04 & 18.88$\pm$0.02 & 18.60$\pm$0.03 & 15.28$\pm$0.04 & 14.76$\pm$0.06 & 4.79  & $-$27.22  & Paper I \\
J081333.33+350810.78 & 20.49$\pm$0.03 & 18.97$\pm$0.02 & 18.94$\pm$0.04 & 15.79$\pm$0.05 & 15.06$\pm$0.08 & 4.92  & $-$27.27  & DR12 \\
J082454.02+130216.98 & 21.36$\pm$0.06 & 19.90$\pm$0.03 & 19.43$\pm$0.07 & 16.43$\pm$0.08 & 15.86$\pm$0.18 & 5.15  & $-$27.03  & DR12 \\
J083832.31$-$044017.47 & 21.20$\pm$0.06 & 19.62$\pm$0.03 & 19.21$\pm$0.07 & 15.58$\pm$0.04 & 15.06$\pm$0.08 & 4.75  & $-$27.01  & Paper I \\
J084631.53+241108.37 & 20.76$\pm$0.03 & 19.15$\pm$0.02 & 19.27$\pm$0.04 & 15.66$\pm$0.05 & 15.14$\pm$0.12 & 4.73  & $-$26.80  & DR12 \\
J085430.37+205650.84 & 21.99$\pm$0.10 & 19.37$\pm$0.03 & 19.38$\pm$0.06 & 16.21$\pm$0.07 & 15.17$\pm$0.10 & 5.17  & $-$27.01  & DR12 \\
J091543.64+492416.65 & 20.93$\pm$0.05 & 19.57$\pm$0.02 & 19.40$\pm$0.06 & 16.41$\pm$0.07 & 15.54$\pm$0.10 & 5.19  & $-$27.22  & DR12 \\
J094108.36+594725.76 & 20.56$\pm$0.04 & 19.27$\pm$0.02 & 19.28$\pm$0.06 & 16.44$\pm$0.06 & 15.88$\pm$0.13 & 4.86  & $-$26.81  & DR12 \\
J095707.68+061059.55 & 20.60$\pm$0.03 & 19.21$\pm$0.02 & 18.87$\pm$0.04 & 16.18$\pm$0.07 & 15.54$\pm$0.13 & 5.14  & $-$27.51  & DR12 \\
J100416.13+434739.12 & 20.95$\pm$0.05 & 19.38$\pm$0.02 & 19.31$\pm$0.06 & 16.64$\pm$0.08 & 16.11$\pm$0.18 & 4.84  & $-$26.87  & DR12 \\
J102622.88+471907.19 & 20.17$\pm$0.03 & 18.73$\pm$0.01 & 18.62$\pm$0.04 & 15.58$\pm$0.04 & 14.85$\pm$0.06 & 4.93  & $-$27.62  & DR12 \\
J104041.10+162233.87 & 20.50$\pm$0.03 & 18.82$\pm$0.07 & 18.75$\pm$0.04 & 16.14$\pm$0.06 & 15.19$\pm$0.12 & 4.80  & $-$27.19  & DR12 \\
J104242.41+310713.20 & 20.37$\pm$0.04 & 18.98$\pm$0.02 & 18.96$\pm$0.05 & 16.18$\pm$0.06 & 15.59$\pm$0.12 & 4.70  & $-$27.12  & DR12 \\
J104325.56+404849.49 & 20.70$\pm$0.04 & 19.02$\pm$0.02 & 19.09$\pm$0.04 & 15.87$\pm$0.05 & 15.06$\pm$0.08 & 4.91  & $-$27.07  & DR12 \\
J105020.41+262002.33 & 20.74$\pm$0.04 & 19.39$\pm$0.02 & 19.34$\pm$0.06 & 16.38$\pm$0.07 & 15.70$\pm$0.14 & 4.86  & $-$27.09  & DR12 \\
J105123.04+354534.31 & 20.23$\pm$0.03 & 18.42$\pm$0.02 & 18.56$\pm$0.04 & 15.46$\pm$0.04 & 14.84$\pm$0.06 & 4.91  & $-$27.75  & DR12 \\
J105322.99+580412.13 & 21.51$\pm$0.05 & 19.80$\pm$0.02 & 19.49$\pm$0.05 & 16.89$\pm$0.09 & 15.96$\pm$0.13 & 5.27  & $-$27.12  & DR12 \\
J112857.85+575909.84 & 20.89$\pm$0.05 & 19.50$\pm$0.03 & 19.20$\pm$0.06 & 16.64$\pm$0.08 & 15.84$\pm$0.14 & 5.00  & $-$27.16  & DR12 \\
J112956.09$-$014212.44 & 21.94$\pm$0.11 & 19.58$\pm$0.02 & 19.47$\pm$0.07 & 15.11$\pm$0.04 & 14.44$\pm$0.05 & 4.87  & $-$26.96  & DR12 \\
J113246.50+120901.70 & 21.25$\pm$0.08 & 19.68$\pm$0.04 & 19.21$\pm$0.06 & 16.14$\pm$0.07 & 15.37$\pm$0.11 & 5.17  & $-$27.41  & DR7 \\
J114657.79+403708.67 & 20.95$\pm$0.05 & 19.38$\pm$0.03 & 19.25$\pm$0.05 & 16.03$\pm$0.06 & 15.29$\pm$0.09 & 4.98  & $-$26.98  & DR12 \\
J120055.62+181733.01 & 21.25$\pm$0.08 & 19.62$\pm$0.03 & 19.44$\pm$0.08 & 16.55$\pm$0.08 & 15.77$\pm$0.13 & 5.00  & $-$26.82  & DR12 \\
J120441.73$-$002149.63 & 20.74$\pm$0.04 & 19.21$\pm$0.02 & 18.93$\pm$0.04 & 15.94$\pm$0.06 & 15.34$\pm$0.11 & 5.09  & $-$27.34  & DR12 \\
J120829.27+394339.72 & 20.79$\pm$0.06 & 19.04$\pm$0.02 & 19.06$\pm$0.05 & 15.80$\pm$0.05 & 15.09$\pm$0.08 & 4.94  & $-$27.25  & Paper I \\
J120952.73+183147.21 & 21.57$\pm$0.11 & 19.80$\pm$0.04 & 19.44$\pm$0.08 & 16.11$\pm$0.06 & 15.38$\pm$0.10 & 5.15  & $-$27.00  & DR12 \\
J124942.12+334953.85\tablenotemark{c} & 20.48$\pm$0.04 & 19.14$\pm$0.02 & 19.08$\pm$0.05 & 16.23$\pm$0.06 & 15.46$\pm$0.09 & 4.93  & $-$27.19  & Schnider1991 \\
J125353.35+104603.19 & 20.95$\pm$0.04 & 19.37$\pm$0.02 & 19.21$\pm$0.05 & 15.33$\pm$0.04 & 14.75$\pm$0.07 & 4.91  & $-$27.11  & DR12 \\
J131234.08+230716.36 & 20.74$\pm$0.04 & 19.30$\pm$0.02 & 18.97$\pm$0.04 & 15.89$\pm$0.05 & 15.28$\pm$0.08 & 4.89  & $-$27.22  & DR12 \\
J131814.03+341805.64 & 20.59$\pm$0.03 & 19.05$\pm$0.02 & 18.83$\pm$0.04 & 15.20$\pm$0.03 & 14.54$\pm$0.05 & 4.82  & $-$27.32  & DR7 \\
J133257.45+220835.91 & 21.12$\pm$0.04 & 19.26$\pm$0.02 & 19.23$\pm$0.04 & 15.69$\pm$0.05 & 14.89$\pm$0.06 & 5.11  & $-$27.39  & Paper I \\
J134015.03+392630.70 & 21.19$\pm$0.04 & 19.39$\pm$0.02 & 19.19$\pm$0.05 & 16.09$\pm$0.05 & 15.48$\pm$0.08 & 5.03  & $-$27.17  & DR12 \\
J134040.24+281328.16 & 21.91$\pm$0.10 & 20.02$\pm$0.03 & 19.48$\pm$0.08 & 16.13$\pm$0.06 & 15.03$\pm$0.07 & 5.34  & $-$27.20  & DR7 \\
J134154.02+351005.71 & 21.32$\pm$0.05 & 19.68$\pm$0.02 & 19.45$\pm$0.05 & 16.29$\pm$0.06 & 15.64$\pm$0.11 & 5.23  & $-$26.95  & DR12 \\
J134408.62+152125.05 & 21.01$\pm$0.06 & 19.40$\pm$0.02 & 19.37$\pm$0.06 & 16.20$\pm$0.06 & 15.68$\pm$0.13 & 4.87  & $-$27.07  & DR12 \\
J134819.88+181925.82 & 20.80$\pm$0.04 & 19.15$\pm$0.02 & 19.18$\pm$0.05 & 16.13$\pm$0.06 & 15.57$\pm$0.11 & 4.94  & $-$27.10  & DR12 \\
J140404.65+031403.85 & 20.93$\pm$0.06 & 19.52$\pm$0.03 & 19.26$\pm$0.07 & 16.09$\pm$0.05 & 15.38$\pm$0.09 & 4.90  & $-$26.92  & DR12 \\
J141839.99+314244.07 & 21.54$\pm$0.06 & 19.69$\pm$0.03 & 19.27$\pm$0.06 & 15.78$\pm$0.04 & 15.11$\pm$0.07 & 4.85  & $-$26.92  & DR12 \\
J142325.92+130300.71 & 21.17$\pm$0.05 & 19.67$\pm$0.02 & 19.39$\pm$0.08 & 15.98$\pm$0.05 & 15.46$\pm$0.09 & 5.02  & $-$26.93  & DR12 \\
J142526.10+082718.46 & 20.54$\pm$0.03 & 18.77$\pm$0.02 & 18.92$\pm$0.04 & 15.96$\pm$0.04 & 15.41$\pm$0.08 & 4.94  & $-$27.13  & DR12 \\
J142634.33+204336.38 & 20.66$\pm$0.03 & 19.13$\pm$0.02 & 18.84$\pm$0.04 & 15.59$\pm$0.04 & 14.99$\pm$0.06 & 4.82  & $-$27.47  & DR12 \\
J143605.00+213239.25 & 21.55$\pm$0.07 & 19.95$\pm$0.03 & 19.28$\pm$0.06 & 16.42$\pm$0.06 & 15.88$\pm$0.11 & 5.22  & $-$27.11  & DR12 \\
J143704.82+070807.72 & 20.62$\pm$0.04 & 19.17$\pm$0.02 & 19.16$\pm$0.05 & 16.14$\pm$0.06 & 15.62$\pm$0.12 & 4.93  & $-$27.10  & Paper I \\
J143751.83+232313.35 & 21.19$\pm$0.06 & 19.45$\pm$0.02 & 19.16$\pm$0.06 & 15.89$\pm$0.04 & 14.97$\pm$0.06 & 5.31  & $-$27.29  & DR12 \\
J144350.67+362315.14 & 22.35$\pm$0.14 & 20.15$\pm$0.03 & 19.47$\pm$0.06 & 15.90$\pm$0.04 & 14.90$\pm$0.05 & 5.12  & $-$27.21  & DR12 \\
J152302.90+591633.05 & 21.39$\pm$0.06 & 19.54$\pm$0.02 & 19.22$\pm$0.05 & 15.64$\pm$0.03 & 15.13$\pm$0.05 & 5.11  & $-$27.40  & Paper I \\
J153650.26+500810.33 & 20.18$\pm$0.03 & 18.48$\pm$0.02 & 18.51$\pm$0.03 & 15.13$\pm$0.03 & 14.52$\pm$0.04 & 4.93  & $-$27.70  & DR12 \\
J155657.36$-$172107.56 & 19.94$\pm$0.04 & 18.43$\pm$0.02 & 18.43$\pm$0.05 & 15.09$\pm$0.04 & 14.59$\pm$0.06 & 4.75  & $-$27.92  & Paper I \\
J160111.17$-$182835.09 & 20.98$\pm$0.15 & 19.37$\pm$0.05 & 18.89$\pm$0.09 & 15.65$\pm$0.05 & 15.05$\pm$0.08 & 5.06  & $-$27.53  & Paper I \\
J160734.23+160417.44 & 20.53$\pm$0.03 & 19.15$\pm$0.02 & 19.09$\pm$0.06 & 16.09$\pm$0.06 & 15.43$\pm$0.09 & 4.76  & $-$27.08  & DR12 \\
J161622.11+050127.71 & 20.16$\pm$0.03 & 18.67$\pm$0.02 & 18.59$\pm$0.04 & 15.90$\pm$0.06 & 15.17$\pm$0.09 & 4.87  & $-$27.80  & DR7 \\
J162045.64+520246.65 & 20.77$\pm$0.04 & 18.97$\pm$0.02 & 18.94$\pm$0.04 & 15.30$\pm$0.03 & 14.70$\pm$0.04 & 4.79  & $-$27.31  & Paper I \\
J162315.28+470559.90 & 20.87$\pm$0.05 & 19.52$\pm$0.03 & 19.23$\pm$0.07 & 15.57$\pm$0.03 & 14.76$\pm$0.05 & 5.13  & $-$27.62  & Paper I \\
J162623.38+484136.47 & 20.06$\pm$0.02 & 18.50$\pm$0.01 & 18.40$\pm$0.03 & 15.51$\pm$0.04 & 15.01$\pm$0.05 & 4.84  & $-$27.86  & DR12 \\
J162626.50+275132.50 & 21.47$\pm$0.06 & 19.17$\pm$0.02 & 18.53$\pm$0.03 & 14.97$\pm$0.03 & 14.21$\pm$0.04 & 5.16  & $-$27.95  & DR12 \\
J162838.84+063859.15 & 20.88$\pm$0.04 & 19.56$\pm$0.02 & 19.40$\pm$0.05 & 16.68$\pm$0.09 & 15.93$\pm$0.17 & 4.85  & $-$27.05  & Paper I \\
J163810.39+150058.26 & 20.53$\pm$0.04 & 18.83$\pm$0.02 & 18.53$\pm$0.04 & 15.10$\pm$0.04 & 14.53$\pm$0.05 & 4.76  & $-$27.57  & Paper I \\
J165354.62+405402.21 & 20.50$\pm$0.03 & 18.59$\pm$0.01 & 18.86$\pm$0.05 & 15.42$\pm$0.03 & 14.72$\pm$0.05 & 4.96  & $-$27.39  & DR12 \\
J165436.85+222733.80 & 19.74$\pm$0.02 & 18.17$\pm$0.01 & 18.08$\pm$0.03 & 15.14$\pm$0.04 & 14.58$\pm$0.05 & 4.70  & $-$27.99  & DR12 \\
J165635.46+454113.55 & 21.51$\pm$0.06 & 19.70$\pm$0.02 & 19.06$\pm$0.04 & 16.22$\pm$0.28 & 15.53$\pm$0.07 & 5.34  & $-$27.64  & Paper I \\
J165902.12+270935.19 & 20.95$\pm$0.07 & 19.34$\pm$0.03 & 18.70$\pm$0.04 & 15.92$\pm$0.05 & 15.14$\pm$0.07 & 5.31  & $-$27.92  & DR7 \\
J173744.87+582829.66 & 20.79$\pm$0.05 & 19.27$\pm$0.02 & 19.15$\pm$0.06 & 16.17$\pm$0.05 & 15.56$\pm$0.07 & 4.92  & $-$27.30  & DR7 \\
J175114.57+595941.47 & 20.75$\pm$0.04 & 19.09$\pm$0.02 & 18.78$\pm$0.04 & 15.66$\pm$0.03 & 15.09$\pm$0.05 & 4.83  & $-$27.24  & Paper I \\
J175244.10+503633.05 & 20.85$\pm$0.04 & 18.82$\pm$0.02 & 18.87$\pm$0.05 & 15.13$\pm$0.03 & 14.40$\pm$0.03 & 5.02  & $-$27.50  & Paper I \\
J211105.62$-$015604.14 & 19.78$\pm$0.02 & 18.11$\pm$0.02 & 18.14$\pm$0.03 & 15.02$\pm$0.04 & 14.41$\pm$0.05 & 4.85  & $-$28.21  & Paper I \\
J215216.10+104052.44 & 19.97$\pm$0.03 & 18.36$\pm$0.02 & 18.22$\pm$0.03 & 14.67$\pm$0.03 & 14.02$\pm$0.04 & 4.79  & $-$28.03  & Paper I \\
J220008.67+001744.93 & 20.68$\pm$0.04 & 19.09$\pm$0.02 & 19.29$\pm$0.06 & 16.20$\pm$0.07 & 15.48$\pm$0.13 & 4.77  & $-$26.93  & DR12 \\
J220106.63+030207.71 & 20.58$\pm$0.03 & 19.11$\pm$0.02 & 18.90$\pm$0.04 & 15.98$\pm$0.06 & 15.20$\pm$0.10 & 5.06  & $-$27.59  & Paper I \\
J220226.77+150952.38 & 20.28$\pm$0.03 & 18.69$\pm$0.02 & 18.47$\pm$0.03 & 15.74$\pm$0.05 & 15.20$\pm$0.08 & 5.07  & $-$28.02  & Paper I \\
J222509.19$-$001406.82 & 20.46$\pm$0.04 & 19.01$\pm$0.03 & 18.71$\pm$0.04 & 15.93$\pm$0.05 & 15.41$\pm$0.11 & 4.85  & $-$27.37  & DR12 \\
J222514.38+033012.50 & 21.74$\pm$0.14 & 20.02$\pm$0.05 & 19.47$\pm$0.10 & 16.50$\pm$0.08 & 15.69$\pm$0.13 & 5.24  & $-$27.17  & Paper I \\
J222612.41$-$061807.29 & 20.32$\pm$0.04 & 18.76$\pm$0.02 & 18.73$\pm$0.05 & 15.64$\pm$0.05 & 14.96$\pm$0.09 & 5.08  & $-$27.83  & Paper I \\
J225257.46+204625.22 & 20.65$\pm$0.04 & 19.16$\pm$0.02 & 19.23$\pm$0.06 & 16.27$\pm$0.06 & 15.52$\pm$0.10 & 4.91  & $-$27.00  & Paper I \\
J232939.30+300350.78 & 20.87$\pm$0.05 & 19.37$\pm$0.02 & 18.93$\pm$0.04 & 16.21$\pm$0.06 & 15.43$\pm$0.10 & 5.24  & $-$27.72  & Paper I \\
J234241.13+434047.46 & 21.17$\pm$0.06 & 19.26$\pm$0.02 & 18.97$\pm$0.05 & 15.57$\pm$0.04 & 14.73$\pm$0.06 & 4.99  & $-$26.93  & Paper I \\
J234433.50+165316.48 & 20.23$\pm$0.03 & 18.46$\pm$0.02 & 18.52$\pm$0.03 & 15.22$\pm$0.04 & 14.56$\pm$0.06 & 5.00  & $-$27.93  & Paper I 
   \enddata
   \tablenotetext{a}{Quasars from the SDSS DR7 and DR12 quasar catalogs are labeled as 'DR7' and 'DR12'. Quasars newly identified by us are labeled as 'Paper I'. Two quasars from \cite{mcgreer09} and \cite{schneider91} were also spectroscopically observed by us and we use our new spectra to do the $M_{1450}$ calculation. See the details for our newly observed quasars in Paper I. All $M_{1450}$ values are corrected using our adopted cosmology.}
   \tablenotetext{b}{This quasar discovered by \cite{mcgreer09} using radio-selection method was also observed by us and we use the new spectra for the $M_{1450}$ calculation.}
   \tablenotetext{c}{This quasar discovered by \cite{schneider91} was also observed by us and we use the new spectra for the $M_{1450}$ calculation.}
\end{deluxetable*}

\begin{figure}
\includegraphics[width=0.45\textwidth]{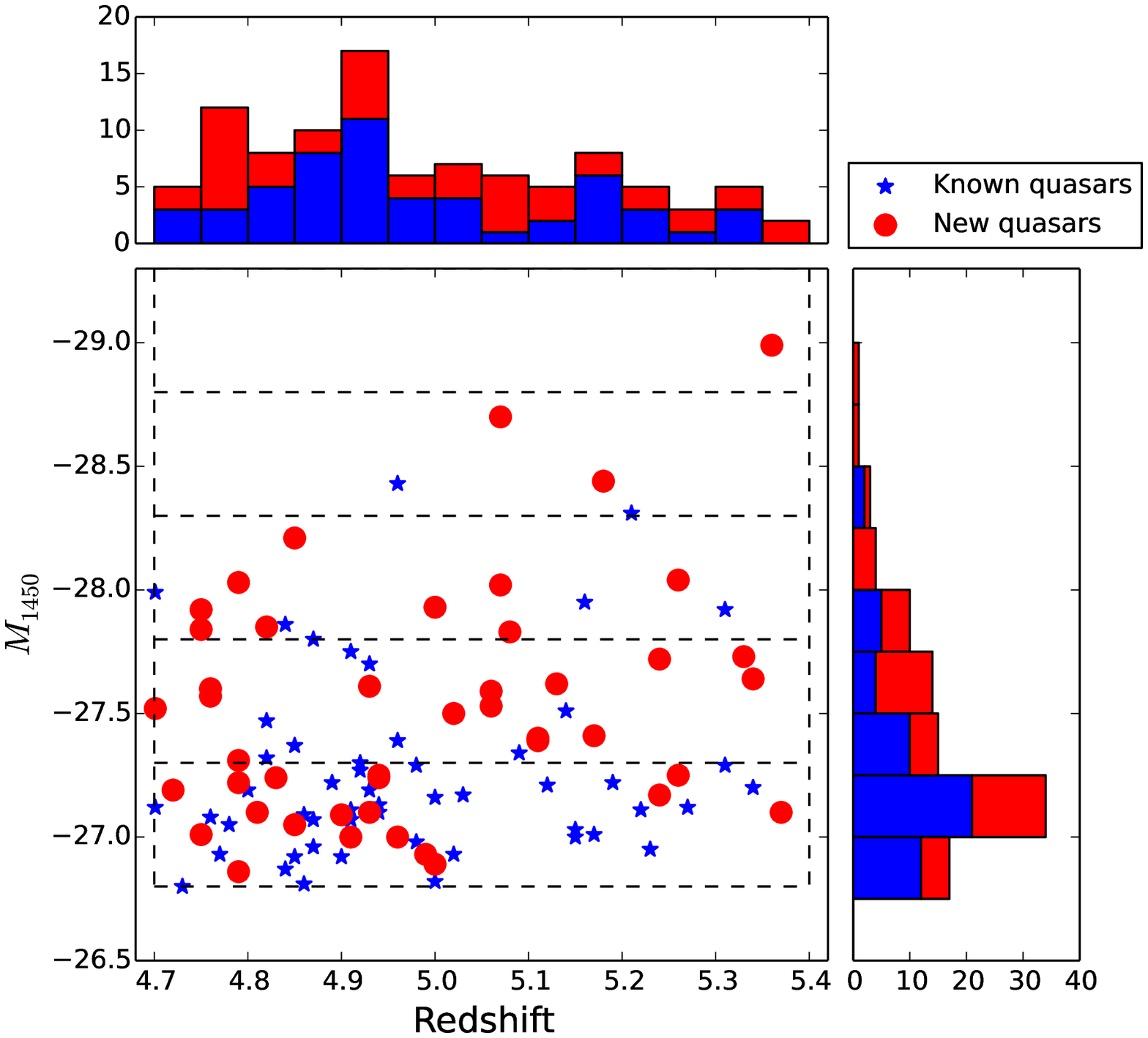}
\centering
\caption{The redshift and $M_{1450}$ distributions of both our newly identified luminous quasars (red dots) and known quasars (blue stars) at $4.7\lesssim z \lesssim 5.4$ and $-29.3 \lesssim M_{1450} \lesssim $-$26.8$. The known quasars are from the SDSS DR7 and DR12 quasar catalogs \citep[][P{\~a}ris et al. in prep]{schneider10}, \cite{mcgreer09} and \cite{schneider91}. This is the sample used for QLF measurement. The dashed lines represent the redshift and magnitude bins for determining the binned QLF. We use only one redshift bin. }
\label{fig1}
\end{figure}
  
\section{Survey completeness} 
 In this section, we will discuss the incompleteness corrections. We limit our candidates to be brighter than 19.5 mag in the SDSS $z$ band, the ALLWISE detection is not deep enough for all quasars in this magnitude range and its depth highly depends on sky position. Therefore, we first model the ALLWISE incompleteness caused by survey depth. We then correct for the incompleteness of our SDSS-ALLWISE color-color selection. Besides, there are 11 candidates that have not been observed, which leads to an incompleteness. For photometric completeness, we visually inspected images of each candidate. To see how many quasars will be missed in this step, we randomly selected 2000 SDSS images of point sources in the same magnitude range as our quasar candidates. We divided them into four groups and visually inspected images. The fraction of rejected images is 2\% - 4\% in each group. It is difficult to obtain a more accurate value of this incompleteness and this effect is much smaller than the error of QLF, thus image selection is not included in our incompleteness correction.

\subsection{Model ALLWISE Incompleteness}
The magnitude limit of our main sample is SDSS $z =$ 19.5, which is much brighter than the flux limit (5 $\sigma$) of the SDSS survey. Therefore, within our magnitude limit, the SDSS detections can be considered as complete. Our survey adds ALLWISE W1 and W2 photometric data into the selection, and thus we need to consider the detection incompleteness caused by the shallower ALLWISE detection limit. We correct this by using the ALLWISE detection completeness from the Explanatory Supplement to the AllWISE Data Release Products\footnote{http://wise2.ipac.caltech.edu/docs/release/allwise/}, which is a function of frames coverage and flux in W1 and W2 bands respectively. Figure 2 represents the empirical models of 2D detection completeness in W1 and W2 bands. As shown, our sample limited with W1 $<$ 17 and $\sigma_{W2} <$ 0.2 will be effected slightly by the detection incompleteness at the faint end.

\begin{figure}
\includegraphics[width=0.55\textwidth]{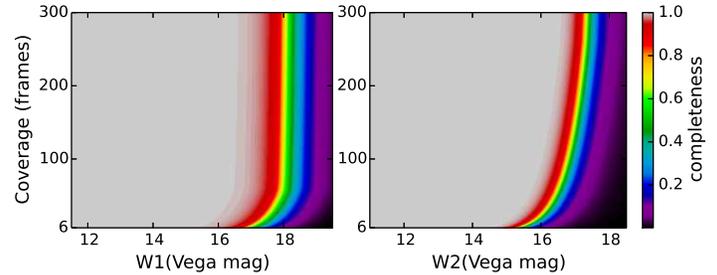}
\centering
\caption{The empirical models of 2D detection completeness in ALLWISE W1 and W2 bands, which is from the Explanatory Supplement to the AllWISE Data Release Products. Our sample limited with W1 $<$ 17 and $\sigma_{W2} <$ 0.2 will be effected slightly by the detection incompleteness at the faint end.}
\label{fig4}
\end{figure}
  
The ALLWISE coverage depends on the sky position. To take the position-dependence into account on completeness correction, we mapped the ALLWISE spatial surveying depth within SDSS footprint. We firstly randomly generated $\sim$1220000 positions in the whole SDSS footprint and derived the ALLWISE coverage map in SDSS footprint by matching($1'$) positions to the nearest ALLWISE sources. A detection with coverage $\le$ 5 could be contaminated by random pixel variations such as cosmic rays because they are at or below the threshold for ALLWISE statistically viable outlier detection and rejection. So we removed all positions with frame coverage $\le$ 5. 
There are only 269 (0.02\%) positions with coverage $\le$ 5. Figure 3 shows the distributions of W1/W2 frame coverages in the whole SDSS footprint. The average coverage is 36 in both W1 and W2 bands. The 10\% and 90\% tile coverage in W1/W2 band are 23/22 and 56/56. We use our ALLWISE coverage map to correct the detection incompleteness (See details in Sec. 3.2).

\begin{figure}
\includegraphics[width=0.55\textwidth]{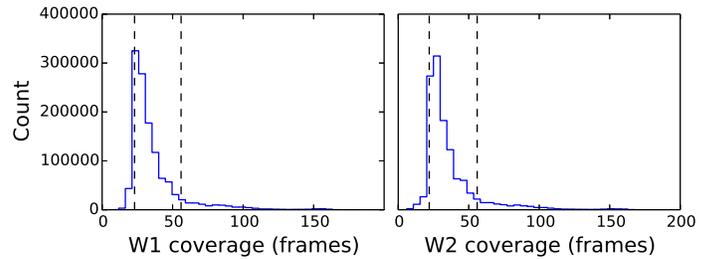}
\centering
\caption{The distributions of W1/W2 frame coverages in the whole SDSS footprint. Dashed lines show the 10\% and 90\% tile ALLWISE coverage. }
\label{fig4}
\end{figure}

ALLWISE coverage also affects the photometric errors of detected sources. The photometric error in W1/W2 will be a function of magnitude and coverage. We used all point sources in our ALLWISE sources sample discussed above to model the empirical magnitude-coverage-magnitude error relations for the ALLWISE W1 and W2 bands. The ALLWISE sensitivity improves approximately as the square root of the depth of coverage, $\sigma \propto$ 1/$\sqrt{N_{cov}}$, $N_{cov}$ is the number of frame coverage. Considering this, we first eliminated the effect of coverage on magnitude errors and then fit the relations between W1/W2 magnitude and coverage-corrected magnitude error. Based on the WISE all-sky magnitude-error relation \citep{wright10}, the final ALLWISE magnitude-coverage-error relation we obtained is 

\begin{equation}
  \sigma(m, N_{cov}) =  a + [2.5/ \ln(10)]n/10^{-0.4m}/\sqrt{N_{cov}} ~.
\end{equation} 
Where m is the magnitude in W1/W2. Constant $a$ is basic photometric error, equals to 0.01 in WLSE all-sky photometry \cite{wright10}. We found that $a$ should be 0.022 for W1 and 0.019 for W2 in ALLWISE photometry. The best-fitted parameter $n$ is 6.43e-8 for W1, 2.71e-7 for W2. Figure 4 shows our empirical model compared with observed data.

\begin{figure}
\includegraphics[width=0.5\textwidth]{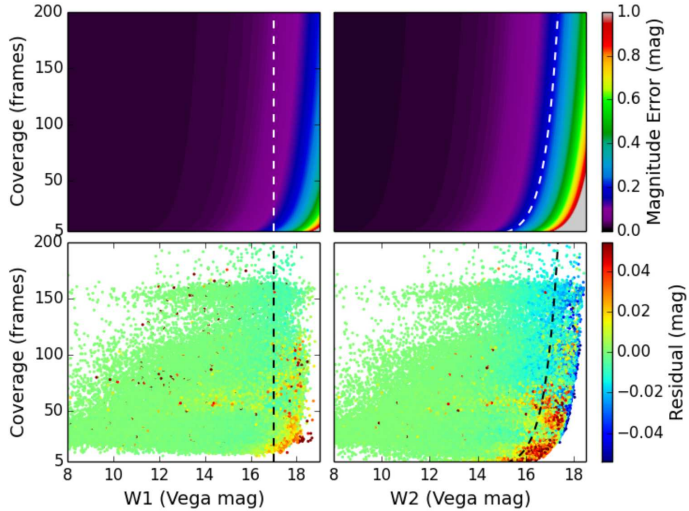}
\caption{{\bf upper:} The magnitude-coverage-magnitude error relations for the ALLWISE W1 and W2 bands. {\bf Bottom:} The residuals of magnitude error, observed data minus model fitted. The daseed lines denote our W1\&W2 magnitude limit (W1 $<$ 17, $\sigma_{W2} <$ 0.2). Our model reproduced the ALLWISE photometric errors in SDSS footprint well. }
\label{fig2}
\end{figure}

\subsection{Selection function of color-color selection}
  To estimate the completeness of our selection criteria, we generate a sample of simulated quasars following the procedure in \cite{fan99a}. M13 updated the spectral model of \cite{fan99a} and applied it to higher redshift, assuming that the quasar spectral energy distributions (SEDs) do not evolve with redshift \citep{kuhn01, yip04, jiang06}. We extend this model toward redder wavelengths to cover the ALLWISE W1, W2 bands for quasars at $z$  = 4 to 6 (McGreer et al. in prep.). The quasar spectrum from M13 is modeled as a power law continuum with a break at 1100\AA. For redder wavelength coverage, we added three new breaks at 5700\AA, 10850\AA, 22300\AA. The slope ($\alpha_{\nu}$) from 5700$\rm \AA$ to 10850$\rm \AA$ follows a Gaussian distribution of $\mu$($\alpha$) = $-$0.48 and $\sigma$($\alpha$) = 0.3; the middle range has a slope with the distribution of $\mu$($\alpha$) = $-$1.74 and $\sigma$($\alpha$) = 0.3; and at the red end, the slope distribution has $\mu$($\alpha$) = $-$1.17 and $\sigma$($\alpha$) = 0.3 \citep{glikman06}. The parameters of emission lines  are derived from the composite quasar spectra \citep{glikman06}. Although the composite spectrum from Glikmann et al. 2006 is constructed from fainter lower redshift quasars, it does not have obvious difference with composite spectra built from luminous quasars at high redshift \cite[e.g.][]{selsing16}. It is the only one we know, which can cover both W1 and W2 bands in the redshift range of our simulation (4 $< z <$ 6). The IGM absorption model is the same as M13, which extend the Ly$_{\alpha}$ forest model based on the work of \cite{WP11} to higher redshift by using the observed number densities of high column density systems \citep{SC10}. Compared to M13, we have made minor modifications for Fe emission. We use the template from \cite{vestergaard01} for wavelengths shorter than 2200\AA. For 2200-3500\AA, we use the template from \cite{tsuzuki06} which separates the FeII emission from the MgII $\lambda$2798 line. A template from \cite{boroson92} covering 3500-7500$\rm \AA$ is also added.
  
Based on this model, we generate a sample of simulated quasars and then calculate the selection function of our color-color selection. We construct a grid of quasars in the redshift range 4 $< z <$ 6 and the luminosity range $-29.5 < M_{1450} < -25.5$. A total of 314,000 simulated quasars has been generated and evenly distributed in the ($M_{1450}$, $z$) space. There are $\sim$ 200 quasars in each ($M_{1450}$, $z$) bin with $\Delta M$= 0.1 and $\Delta z$= 0.05. We assign optical photometric errors, which are from the SDSS main survey, and photometric uncertainties of the W1 \& W2 bands using the empirical magnitude-coverage-error relations discussed above. We added the ALLWISE detection completeness into the selection probability calculation. 

We calculate the ALLWISE detection probability by randomly choosing a unique sky position from our 1220000 positions for each simulated quasar, and thus obtained a ALLWISE detection probability of each simulated quasar based on its frames coverage and W1 \& W2 magnitude. For each ($M_{1450}$, $z$) bin ($\Delta M$= 0.1 and $\Delta z$= 0.05) discussed above, we obtain a mean detection probability. Then we calculate the fraction of simulated quasars selected by our selection criteria in each ($M_{1450}$, $z$) bin as the selection probability, shown in Figure 5.

As shown in Figure 5, after relaxing the traditional $r-i$/$i-z$ color cut and adding the W1-W2 color, our color selection criteria show a high completeness at $4.8 < z < 5.2$ and extend the selection region to z $\sim$ 5.4. Within the central bright region ($4.8 < z< 5.2$ and $M_{1450} < -26.8$), the mean completeness reaches 78\%. Extended to the range of $4.7 < z < 5.4$, the mean completeness is $\sim$ 60\%. At redshift lower than 4.5 or higher than 5.4, the completeness is below than 5\%. At $z < 4.7$, the W1-W2 color becomes bluer; our W1-W2 $>$ 0.5 cut will miss some quasars at $z <$ 4.7 (See Fig. 3 in Paper I ). However, the exact completeness is more sensitive to the assumption we made about the rest-frame optical continuum of high-redshift quasars at $4.5 < z < 4.7$ due to the fast change of W1-W2 color here. The uncertainty of simulation in this redshift range is higher than that at $z > 4.7$. Therefore, we restrict our quasar sample for QLF calculation to $z\ge 4.7$. At 5.2 $< z < $ 5.4, although the completeness becomes lower, our selection has explored a higher redshift range with higher completeness than previous works at z $\sim$ 5. Thus, we limit our sample with $z < 5.4$.

To see how the ALLWISE coverage effect our selection function, we also calculate the selection function by assuming fixed number of coverage, 10\% tile ALLWISE coverage ($N_{cov}$ = 23 for W1, $N_{cov}$ = 22 for W2), shown in Figure 6. The difference in selection function between using 10\% tile coverage and position-dependent coverage is less than 5\% at $M_{1450} < -27$ and increases to $\sim$10\% at $M_{1450} < -26.8$, to 15\% - 20\% at fainter range. We also compare the QLF result base on 10\% tile coverage and position-dependent coverage. The change of the parameters of best-fitted QLFs is $\sim$ 0.05, much smaller than error bars. We use the position-dependent coverage selection function to calculate the parametric QLF. When we calculate the selection probability of each quasar in our sample for binned QLF measurement, we use the real ALLWISE coverage to calculate the ALLWISE detection completeness of each quasar. The agreement between binned QLF and best fit parametric QLF (See section 4.1 \& 4.2) shows that our ALLWISE coverage model and the selection function using mean detection incompleteness are reasonable.

\begin{figure}
\includegraphics[width=0.5\textwidth]{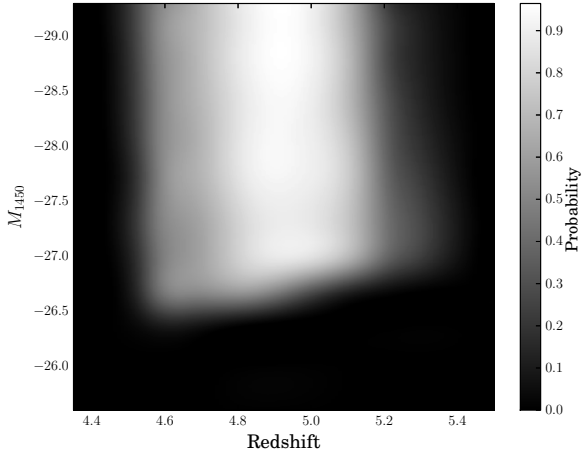}
\caption{The selection function of our survey ($z_{SDSS} < 19.5$) based on position-dependent ALLWISE coverage. The probability is the fraction of simulated quasars which can be selected by our selection criteria in each ($M_{1450}$,z) bin. }
\label{fig3}
\end{figure}

\begin{figure}
\includegraphics[width=0.5\textwidth]{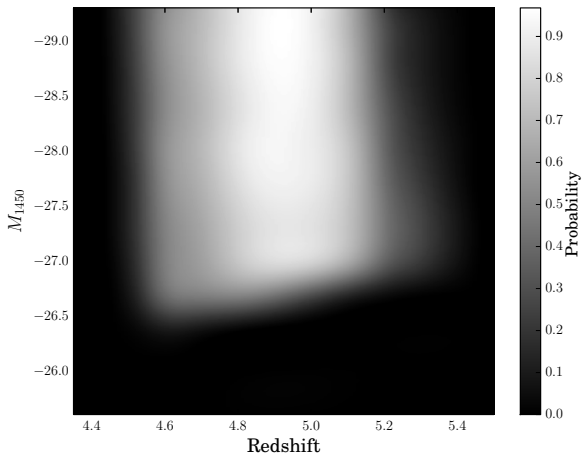}
\caption{The selection function of our survey ($z_{SDSS} < 19.5$) using 10\% tile ALLWISE coverage, $N_{cov}$ = 23 for W1, $N_{cov}$ = 22 for W2. Comparing with Figure 5, the change of probability is small. }
\label{fig3}
\end{figure}
\subsection{Spectroscopic incompleteness}
 
We spectroscopically observed 99 out of 110 candidates. The spectroscopic completeness reaches 100\% at $z_{SDSS} \le$ 19; at the fainter end, the completeness is lower but it still has a high value around 80\%. The histogram of our observed and unobserved candidates is shown in Figure 7. The completeness is a function of $z$ band magnitude. We use this function to correct the incompleteness from spectral coverage, assuming the probability of an unobserved candidate to be a quasar is the same as in the observed sample. As shown, the quasar fraction in our unknown candidate sample becomes lower at the faint end. That is caused by the fact that there are more known quasars at $z_{SDSS} >$ 19, and these known quasars are not plotted in this figure. 

\begin{figure}
\includegraphics[width=0.5\textwidth]{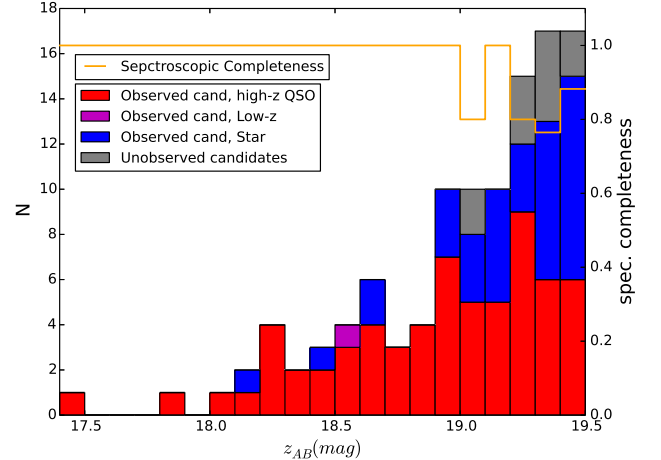}
\caption{Spectroscopic incompleteness of our 110 $z \sim$ 5 quasar candidates. The orange line denotes the spectroscopic incompleteness as a function of $z$  band magnitude. The histogram is divided into several components filled by different colors and represents newly identified high redshift quasars (red), low redshift quasars (purple), stars (blue) and unobserved candidates (grey). }
\label{fig5}
\end{figure}  
  
\section{A new determination of the QLF at $z \sim 5$}
\subsection{Binned QLF}
   To compute the binned QLF, we divide our sample into several bins. Due to the narrow redshift interval of our sample, we only use one redshift bin and do not include any evolution with redshift. We then divide our sample into 5 magnitude bins with $\Delta M_{1450}$ = 0.5 mag over the magnitude range $-26.8 < M_{1450} < $-$29.3$ (See Fig. 1). We calculate the binned luminosity function by using the \cite{page00} modification of the 1/V$_{\rm a}$ method \citep{schmidt68, avni80} for flux limit correction. The final selection function is applied after all incompleteness corrections have been applied for each quasar. The result for the binned QLF and number counts are listed in Table 2. In the table, the number counts and  corrected number counts derived by applying all incompleteness corrections are denoted as $N$ and $N_{cor}$ respectively. The result is also displayed in Figure 8 as red squares together with the binned QLF data from the SDSS main (black) and Stripe 82 (blue) samples in M13 for comparison. Data from M13 have been corrected to z=5.05 by using the quasar redshift evolution at high redshifts according to \cite{fan01b}. Compared to previous results, our binned QLF has more luminous quasars and extends the measurement of the $z \sim$ 5 QLF to $M_{1450} = -29$, and thus gives a smaller error bar in each bin at $M_{1450} <$ $-$27.05. Our data show a similar result, but suggest a higher value at the bright end. The binned QLF in the brightest bin has a large error bar due to the fact that there is only one quasar in this bin.

\begin{deluxetable}{ c c c c c c }
\tabletypesize{\scriptsize}
\tablecaption{Binned QLF. \label{tbl-2}}
\tablewidth{0pt}
\tablehead{
  \colhead{$M_{1450}$} &
  \colhead{$N$} &
  \colhead{$N_{cor}$} &
  \colhead{log$\Phi$} &
  \colhead{$\Delta\Phi$ \tablenotemark{a}} 
 }
\startdata
  $-$28.99\tablenotemark{b} & 1 & 18.2 & $-$9.48 & 0.33 \\
 $-$28.55 & 4 & 7.7 & $-$9.86 & 0.08 \\
  $-$28.05 & 14 & 24.4 & $-$9.36 & 0.15 \\
  $-$27.55 & 26 & 44.4 & $-$9.09 & 0.19 \\
  $-$27.05 & 54 & 103.1 & $-$8.70 & 0.32 
\enddata
\tablenotetext{a}{$\Delta\Phi$ is in units of $\rm 10^{-9} Mpc^{-3} mag^{-1}$. }
\tablenotetext{b}{Within the brightest magnitude bin, there is only one quasar. Therefore, we use its $M_{1450}$ as the $M_{1450}$ of this bin.}
\end{deluxetable}

\begin{figure}
\includegraphics[width=0.5\textwidth]{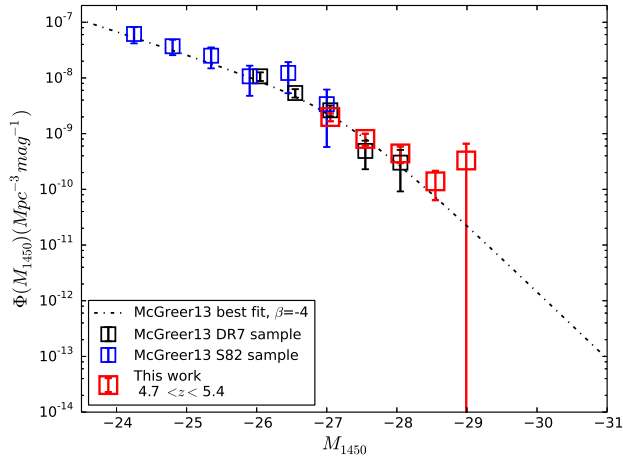}
\caption{Binned QLF at $z$ = 5.05 ($4.7 \le  z < 5.4$). The red squares represent our binned QLF data. The blue and black squares denote the binned QLF data from the Stripe 82 sample and the SDSS main sample in M13. These data have been corrected to $z$ = 5.05 by using the redshift evolution proposed by \cite{fan01b}. The black dash-dotted line shows the best fitting QLF from M13 with the bright end slope $\beta$ = $-$4.}
\label{fig6}
\end{figure}
                   
\subsection{Maximum likelihood fitting}
   The binned QLF result, while non-parametric, is dependent on the choice of binning. Here we derive a parametric QLF by performing a maximum likelihood fit for each quasar in the sample. We model the QLF using the most common double power law form \citep{boyle00}:
 \begin{equation}
	\Phi(M,z) = \frac{\Phi^*(z)}
	     {10^{0.4(\alpha+1)(M-M^*)} + 10^{0.4(\beta+1)(M-M^*)}} ~,
\end{equation}   
where $\alpha$ and $\beta$ are the faint end and the bright end slopes; $M^{*}$ is the break magnitude and $\Phi^*(z)$ is the normalization. These four parameters have been suggested to evolve with redshift. Following previous work, we adopt the rapid decline in quasar number density at high redshift from \cite{fan01b} as the QLF evolution within our narrow redshift interval, $\Phi^*(z) = \Phi^*(z=6)\times10^{k(z-6)}$, where $k=-0.47$ \citep{fan01b}\footnote{To see how the results depend on $\Phi^*(z)$ evolution, we varied the value of k from $-0.3$ to $-0.7$. We find that the form of $\Phi^*(z)$ evolution has little effect on the other parameters when doing the fits. The changes are within $1\sigma$. And due to the narrow redshift range we used, the log($\Phi^*(z)$) is also affected only slightly by varying the form of the evolution.}. Here we also normalize $\Phi^*$ to $z=6$ for easier comparison to the higher redshift results. 

Due to the fact that our quasar sample covers the magnitude range $M_{1450} \le $ $-$26.8, and the break magnitude given by M13 is around $-$26 to $-$27, our sample cannot be used to constrain the faint end slope. For measurement of the break magnitude $M^{*}$, we combine our luminous quasar sample with the S82 and DR7 quasar samples from M13 and then carry out parametric fits for the QLF for all observed quasars in the combined sample. The DR7 sample has a large number of overlaps with our luminous quasar sample. Therefore, we select DR7 quasars only in the magnitude range  $-26.8 < M_{1450} < -25.8$ to construct the combined sample. For the S82 and DR7 samples, we use the same incompleteness corrections as in M13. We use maximum likelihood estimation to derive the fit. The maximum likelihood fit \citep{marshall83} for a luminosity function aims to minimize the log likelihood function S which is equal to $-$2~$\ln$L, where L is the likelihood function:

\begin{equation}
	S = -2 \sum\limits_i^N \ln[\Phi(M_i,z_i)] 
	      + 2\int\int\Phi(M,z)p(M,z)\frac{dV}{dz} dM dz ~,
\end{equation}
where the first term is the sum over all observed quasars in the sample, and the second term is integrated over the full range of absolute magnitude and redshift of the sample \citep{marshall83, fan01a}; $p(M, z)$ is the probability for a quasar to be observed by the survey at given absolute magnitude $M_{1450}$ and redshift z. It includes all incompleteness corrections discussed above. The second term represents the total number of expected quasars in the survey with a given luminosity function, and provides the normalization for the likelihood function. The confidence intervals are determined from the likelihood function by assuming a $\chi^{2}$ distribution of $\Delta$S (= S $-$ $S_{min}$) \citep{lampton76}.

We first fix the faint end slope $\alpha$ to be $-$2.03, as given by M13. We find the luminosity function parameters to be log$\Phi^*(z=6)$= $-$8.82$\pm$0.15, $M_{1450}^{*}$ = $-$26.98$\pm$0.23 and $\beta$ = $-$3.58$\pm$0.24. This result is plotted in Figure 9 and shows excellent agreement with our binned QLF. In order to investigate how the different values of $\alpha$ affect our result, we also assume the faint end slope $\alpha$ to be $-$1.8, similar to what was measured from quasar samples at $z \ge$ 4 \citep{willott10b, glikman10, masters12, mcgreer13}, and to $-$1.5, typical for lower redshift measurements at $z \lesssim$ 3 \citep{croom09}. The values of parameters assuming different faint end slopes are listed in Table 3. When we change the faint end slope $\alpha$ from $-$2.03 to $-$1.8 and $-$1.5, the bright end slope $\beta$ is flattened but only at the $\lesssim 1\sigma$ level. The break magnitude becomes fainter more significantly following the change of $\beta$. If we allow all four parameters to be unconstrained, we derive a steeper bright end slope $\beta = -3.80$ and a very bright break magnitude of $M^{*}_{1450} = -27.33$ with significantly larger error bars. Allowing all parameters to be free has many degeneracies, which is why the uncertainty ranges are larger. In this case, the faint end slope $\alpha$ also becomes steeper ($\alpha = -2.15$). We need more data to better constrain a 4-parameter fit, especially for $M_{1450} < -28.3$. Considering that $\alpha = -2.03$ derived by M13 is a strong constraint on the faint end slope, we adopt the result based on fixed $\alpha = -2.03$ as our best fit. The fitted QLFs for different cases are plotted in Figure 9.

\begin{deluxetable}{ c c c c }
\tabletypesize{\scriptsize}
\tablecaption{Parameters of fits\label{tbl-3}}
\tablewidth{0pt}
\tablehead{
   \colhead{$\alpha$} &
   \colhead{$\beta$} &
   \colhead{$M^{*}_{1450}$} &
   \colhead{log$\Phi^{*}$($z$=6)}
}
\startdata
  $-$2.03 & $-$3.58$\pm$0.24 & $-$26.98$\pm$0.23 & $-$8.82$\pm$0.15\\
  $-$1.80 & $-$3.26$\pm$0.18 & $-$26.28$\pm$0.29 & $-$8.35$\pm$0.17\\
  $-$1.50 & $-$3.03$\pm$0.12 & $-$25.56$\pm$0.29 & $-$7.94$\pm$0.15\\
  $-$2.14$\pm$0.16 & $-$3.80$\pm$0.47& $-$27.32$\pm$0.53 & $-$9.07$\pm$0.40
\enddata
\tablecomments{We fix the faint slope $\alpha$ to be $-$2.03, $-$1.8 and $-$1.5 respectively. And then we allow all four parameters to be free. $\alpha$ = $-$2.03 is measured from the combination of the SDSS S82 and DR7 samples in M13. We adopt the result with fixed $\alpha$= $-$2.03 as our best fit.}
\end{deluxetable}

\begin{figure}
\includegraphics[width=0.5\textwidth]{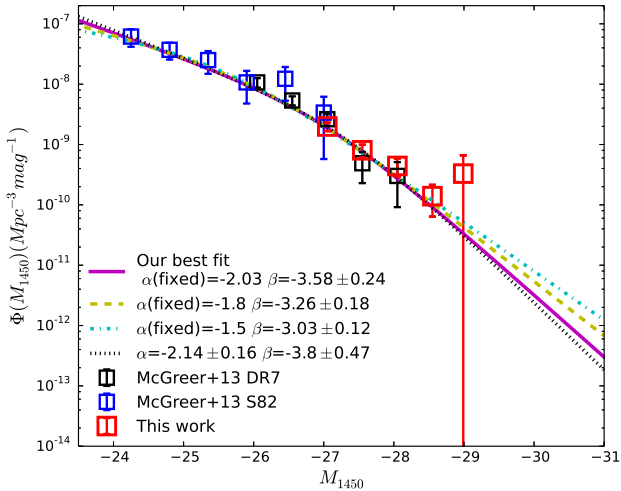}
\caption{Double power law fits using maximum likelihood fitting compared with the binned QLF data from the S82 sample, DR7 sample and our luminous quasar sample. The results based on fixed $\alpha$ and on four free parameters are plotted for comparison. We fix the faint end slope $\alpha$ at $-$2.03 (purple line), $-$1.8 (yellow dashed line) and $-$1.5 (cyan dot-dashed line) and do the fits respectively. Then we also allow all four parameters to be free (Green dashed line). When we change the faint end slope $\alpha$ from $-$2.03 to-1.8 and $-$1.5, the bright end slope $\beta$ is flattened, but it only changes a little. The break magnitude becomes fainter following the change of $\beta$. When we allow all four parameters to be free, we get a steeper bright end slope $\beta$ = $-$3.80 }
\label{fig7}
\end{figure}

\begin{figure}
\includegraphics[width=0.5\textwidth]{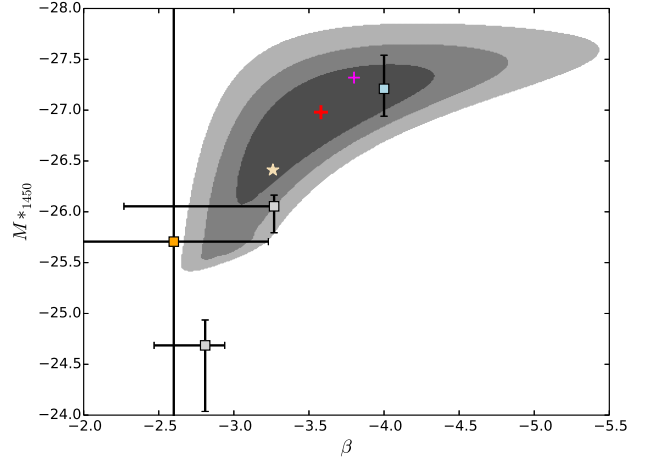}
\caption{Confidence region for $\beta$ and $M^{*}_{1450}$. The regions filled with different shades of grey denote 1-$\sigma$ (68.3\%), 2-$\sigma$ (95.4\%) and 3-$\sigma$ (99.7\%) regions, respectively. For comparison, we plot our best fit result (red cross) and the result of the 4-parameter fit (magenta cross) together with the best fit from other work. The light blue square denotes the best fit from M13 at $z \sim$ 5. The white star shows the result form \cite{willott10b} for fixed $\alpha$ = $-$1.8 (uncertainties of the fit were not reported) at $z \sim$ 6. The yellow square represents the best fit from Masters et al.(2012), which used the faint quasar sample in the COSMOS field in conjunction with the bright quasar sample from \cite{richards06} to model a double power law QLF at $z\sim$ 4. We also plot the points (grey squares) to show the best fits for binned data in the $z$ = 2.2 (left) and 3.4 (right) redshift bins from the BOSS S82 sample (\cite{ross13}). All data have been corrected to our adopted cosmology.}
\label{fig8}
\end{figure}

We calculate the confidence regions to investigate  the degeneracy between the bright end slope and break magnitude; the results are shown in Figure 10. The regions filled with different colors illustrate 1-$\sigma$ (68.3\%), 2-$\sigma$ (95.4\%) and 3-$\sigma$ (99.7\%) regions, respectively. We generate the probability contours by calculating $S_{min}$ for each ($M_{1450}$, $\beta$) point and allowing log$\Phi(z=6)^*$ to be free at each point with the fixed $\alpha = -2.03$.  Figure 10 shows that our data constrain $\beta$ to the range $-4.83 < \beta < -2.78$ at 95\% confidence; this is flatter than the result from M13, which shows $\beta < -3.1$ at 95\% confidence, although the best fit from M13, $\beta$ = $-$4, lies within our 2$\sigma$ region. 

We compare the parameters of our result with previous work at different redshifts to study the evolution of the QLF. In Figure 11, we plot the evolution of the normalization $\Phi^{*}$, the break magnitude $M^{*}_{1450}$ and the bright end slope $\beta$ with redshift. \cite{ross13} measured the QLF at 2.2 $< z <$ 3.5 using the BOSS DR9 quasar sample and concluded that the QLF can be described well by a luminosity evolution and density evolution (LEDE) model at this redshift range. In this model, the evolutions of normalization and break luminosity with redshift are expressed in a log-linear relation, and slopes of the double power law are fixed. M13 add a point at $z$ = 4.9 and combine the result from \cite{masters12} at $z$ =4 and the result from \cite{willott10b} at $z$ = 6 to modify this model. They found that the slope of the normalization evolution was steeper ($c1 = -0.7$) and the slope of the break magnitude evolution was shallower ($c_2 = -0.55$). Now we add our new measurement at $z$ =5.05 and the point at $z$ = 6 from \cite{kashikawa15} in case 1. The result from \cite{kashikawa15} includes the discovery of new faint $z \sim$ 6 quasars and places stronger constraints on the faint end slope and break magnitude of the $z \sim$ 6 QLF. Then we use all of these points to fit the LEDE model. 

\begin{eqnarray}
  \log[\Phi^{*}(z)]  & = & \log[\Phi^{*}(z=2.2)] + c_{1}(z-2.2)~,
  \label{eq:LEDE_Phistar} \\
            M_{i,2}^{*}(z) & = &M_{i,2}^{*}(z=2.2) + c_{2}(z-2.2)~,
  \label{eq:LEDE_Mstar}
\end{eqnarray} 
where $M_{i,2} \equiv M_i(z=2) = M_{1450} - 1.486$ is the absolute $i$-band magnitude at $z=2$ \citep{richards06}, corresponding to rest-frame $\sim2600$\AA\ in the assumption of a spectral index of $\alpha_\nu=-0.5$. We obtain values of log$\Phi^{*}(z=2.2) = -5.87\pm0.07$ and $c1 = -0.81\pm0.03$; $M_{i,2}^{*}(z=2.2) = -26.68\pm0.15$ and $c2 = -0.50\pm0.08$. Note that the errors of parameters are standard deviation errors of fit. We only use these points without uncertainties to do the fit because the uncertainties of the best fit in \cite{willott10b} are not reported. The real errors should be larger than the fitting errors explored here. Our result is consistent with the LEDE model but prefers a steeper slope of log$\Phi^{*}(z)$ evolution and a flatter slope of the break magnitude evolution. \footnote{Our results also can be compared to Fig. 19 of M13; however, the BOSS data used in M13 were based on a pre-publication analysis of the DR9 sample, and were later updated in \cite{ross13}. Here we use the final version of the BOSS data from that work.}

\begin{figure}
\includegraphics[width=0.45\textwidth]{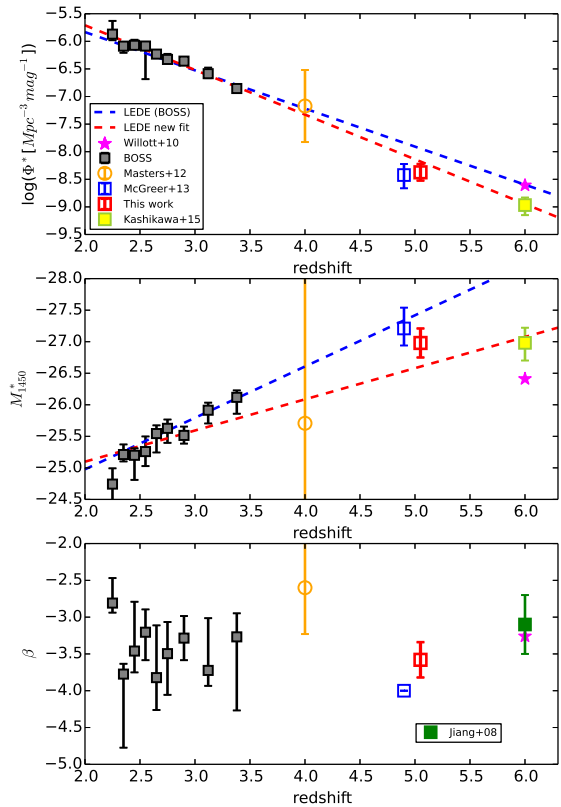}
\caption{Evolution of QLF parameters: normalization $\Phi^{*}$ (upper), break magnitude $M^{*}_{1450}$(middle) and the bright end slope $\beta$ (bottom). We compare our best fit QLF parameters with previous results at similar and different redshifts to show the evolution of parameters from redshift of $z \sim$ 2 to 6. The grey squares at 2.2 $< z <$ 3.5 are the best fits for each redshift bin from the BOSS DR9 QLF, which were measured by using a sample of $\sim$6000 variability selected quasars in Stripe 82 \citep{ross13}. The orange circle is the best fit from \cite{masters12} at $z$ = 4. The blue square represents the best fit from M13 at $z$ = 4.9, and our best fit at $z$ = 5.05 is plotted as a red square. At $z$ = 6, in the $\Phi^{*}$ and $M^{*}_{1450}$ plots, we plot the result from \cite{willott10b} with $\alpha$ = $-$1.8 (magenta star) and \cite{kashikawa15} in their case 1 fits (yellow square). In the $\beta$ $-$ redshift plot, we plot the result from \cite{willott10b} and the single power law fit of \cite{jiang08} at $z$ = 6, because \cite{kashikawa15} fit their data with a fixed bright end slope. The blue dashed lines in the $\Phi^{*}$ and $M^{*}_{1450}$ plots represent the LEDE model from \cite{ross13}. The red dashed lines are our new fits. All data have been corrected to our adopted cosmology. }
\label{fig9}
\end{figure}

\section{Discussion}
\subsection{Contribution to the Ionizing Background}

Previous measurements of the QLF at $z$ = 5 and 6 have shown evidence that quasars cannot produce the entire required ionizing photon background \citep[M13;][]{willott10b, kashikawa15, meiksin05, bolton07} at those redshifts. It is suggested that quasars can contribute $\sim $30\% to 70\% of the ionizing photons required  to maintain full ionization at $z$ = 5 (M13), and produce about several percent to 15\% at $z$ = 6 \citep{willott10b, kashikawa15}, depending on the assumed IGM clumping factor C. Here we update the quasar contribution to the high-redshift ionizing background using our new QLF at $z\sim 5$. 

We calculate the comoving emissivity of quasars at the Lyman limit by $\epsilon(z)=\int\phi(L_{\nu}, z)L_{\nu}dL_{\nu}$ erg s$^{-1}$ Hz$^{-1}$ Mpc$^{-3}$, assuming the escape fraction of ionizing radiation from quasars f = 1. We integrate our parametric QLF $\Phi(M_{1450},z)$ and then convert it into emissivity at $\lambda$ = 912\AA. For the conversion, we adopt the UV slopes from \cite{stevans14}, who suggest a gradual break wavelength at 1000\AA, with the index $\alpha_{\nu} = -1.41$ in the extreme ultraviolet and a spectral index $\alpha_{\nu} = -0.83$ at wavelengths above the break. By integrating our best fit QLF to $M_{1450}$ = $-$20, we derive an ionizing photon density $\dot{N}_Q $ = $ 6.06 \times 10^{49}~\rm{Mpc}^{-3}~{\rm s}^{-1}$. When we use the QLF result with a fixed faint end slope of $\alpha$ = -1.8, the photon number density changes to $\dot{N}_Q $ = $ 4.73 \times 10^{49}~\rm{Mpc}^{-3}~{\rm s}^{-1}$. Using the QLF generated from the 4-parameter fit, we get $\dot{N}_Q $ = $ 7.37 \times 10^{49}~\rm{Mpc}^{-3}~{\rm s}^{-1}$. The change of ionizing photon density is dominated by the change of the break magnitude $M^{*}_{1450}$and the faint end slope $\alpha$. Luminous quasars make little contribution to the ionizing background, so a survey of faint quasars is required to give more accurate measurement.

The required number of photons to balance hydrogen recombination and maintain full ionization was estimated by \cite{madau99} as a function of redshift. The number of required photons at $z$ = 5 is $\dot{N}_{\rm ion} = 3.38\times10^{50} (C/5) ~\rm{Mpc}^{-3}~{\rm s}^{-1}$ in our adopted cosmology. The clumping factor C is crucial to estimate the contribution of quasars to the ionizing background. \cite{madau99} considered a recombination-dominated IGM and suggested a high value for the clumping factor $C=30$. Recent work provides a lower clumping factor $C < 10$. \cite{meiksin05} suggests that $C \approx 5$ and a $C \approx 2\mbox{--}3$ at $z=6$ is suggested by some reionization models \citep{mcquinn11, shull12, finlator12}. For $C = 2$, based on the result from our best fit QLF, quasars are estimated to provide $\sim$ 45\% of the required photons; while for $C = 5$, the fraction changes to 18\%. This result agrees with previous work, suggesting that quasars may play some role in maintaining ionization at $z \sim$ 5 but have low possibility to be the dominant source of ionizing photons (M13). 

\subsection{Radio-loud fraction}
Traditionally, quasars have been divided into two populations, radio-loud and radio-quiet \citep{kellermann89}. The similarity and difference between the evolution of radio-loud and radio-quiet quasars are thought to be related to black hole mass,  accretion and spin.\citep{rees82, wilson95, laor00}. The radio-loud fraction (RLF) has been suggested to evolve with optical luminosity and redshift by some work \citep[e.g.,][]{padovani93, lafranca94, hooper95, jiang07, kratzer15}. In contrast to this, no evolution of the RLF is also found \citep[e.g.,][]{goldschmidt99, stern00, ivezic02, cirasuolo03}. Our luminous quasar sample is selected only by optical and near-infrared colors and thus it can be considered as an unbiased sample for the study of the RLF. 

 We cross match all 99 quasars in our sample with catalogs from Faint Images of the Radio Sky at Twenty-cm \citep[FIRST;][]{becker95} and the NRAO VLA Sky Survey (NVSS) \citep [NVSS;][]{condon98} and find 8 quasars with radio detections. We use 3$''$ matching radius for FIRST data, and 5$''$ for NVSS due to the lower resolution of NVSS. We calculate the radio loudness for these 8 quasars by assuming an optical spectral index of $-$0.5 and radio spectral index of $-$0.5 ($f_{\nu} \propto \nu^{\alpha}$ ) and list them in Table 4. To compare with previous work at higher redshifts \citep{banados15}, we adopt the radio/optical flux density ratio $R_{4400}$ = $f_{\nu,5 GHz}$ /$f_{\nu,4400\AA}$ \citep{kellermann89} and the criterion $R >$ 10 for the definition of a radio loud quasar, where $f_{\nu,5 GHz}$ is the radio flux density at rest-frame 5 GHz, and $f_{\nu,4400\AA}$ is the optical flux density at rest-frame 4400 \AA. There are 7 quasars considered as radio loud quasars among the 99 quasars. Therefore, we find a radio loud fraction RLF $\sim$ 7.1\%. Considering FIRST with its 1mJy flux limit and NVSS with its 2.5 mJy flux limit are not deep enough to detect all radio loud quasars, especially for quasars at $z_{SDSS} > 19$, 7.1\% is a lower limit. This result shows agreement with the result from \cite{banados15} which constrains the RLF at $z \sim$ 6 to be $8.1 ^{+5.0}_{-3.2}$\%. We also do the calculation using a radio spectral index of $-$0.75. The radio loudness based on $\alpha$ =-0.5 to -0.75 increases 12\%-15\%. The radio loud fraction has no change. 
    
 \cite{jiang07} suggest that the RLF is a function of absolute magnitude $M_{2500}$ and redshift at $z <$ 4, and give the best fit for the function. To compare with \cite{jiang07}, we also calculate the radio/optical flux density ratio $R_{2500}$ = $f_{\nu,5 GHz}$ /$f_{\nu,2500\AA}$, where $f_{\nu,2500\AA}$ is the optical flux density at rest-frame 2500 \AA. We convert $M_{1450}$ to $M_{2500}$ by $M_{2500}$ = $M_{1450}$ $-$ 0.3 on the assumption of an optical spectral index $\sim$ $-$0.5. For comparison, here we use the same cosmology as \cite{jiang07} which is $H_{0}$ = 70 km s$^{-1}$ Mpc$^{-1}$, $\Omega_{m}$ = 0.3, and $\Omega_{\Lambda}$ = 0.7. Our sample  covers magnitudes $-$27.03 $< M_{2500} <$ $-$29.22. Due to the fact that there are only 8 quasars with radio detection covering a narrow redshift range, we use only one redshift bin and roughly divide our sample into two magnitude bins, one for  $ M_{2500} <$ $-$28 and the other one for $M_{2500} >$ $-$28. Here we use $R >$ 30 as the definition of radio loud quasar so that FIRST with limiting flux of $\sim$ 1 mJy will be deep enough for our quasar sample. NVSS, with a $\sim$ 2.5 mJy flux limit, is still not deep enough and is able to detect radio-loud quasars ($R >$ 30) at $z \sim$ 5 only down to $M_{2500} \sim$ $-$27.7. Therefore, we calculate the RLF for quasars within the area covered by FIRST and NVSS, respectively. In the bright magnitude bin, there are 13 quasars in the FIRST coverage and 19 quasars in the NVSS coverage. In the faint bin, there are 68 quasars in the FIRST coverage and 80 quasars in the NVSS coverage. The details and radio loudness of radio detected quasars are given in Table 4. We compare our result with the RLF evolution function log (RLF/(1- RLF)) = $-$0.218 $-$  2.096log(1+ z) $-$0.203($M_{2500}$ + 26) from \cite{jiang07}. As shown in Figure 12, our points show an evolution of the RLF with magnitude, but prefer a higher RLF at $z \sim$ 5 than the prediction. Our result suggests that the RLF may evolves with optical luminosity, but it may not decline as rapidly with increasing redshift as measured in \cite{jiang07} at high redshift, although this could be affected by the small number of radio-loud quasars in our sample.
 
\begin{deluxetable*}{l r r r r r r r r r r r r}
\tabletypesize{\tiny}
\tablecaption{Radio detection and radio loudness\label{tbl-4}}
\tablewidth{0pt}
\tablehead{
  \colhead{name} &
  \colhead{$z$} &
  \colhead{m$_{1450}$} &
  \colhead{$f_{1.4GHz, FIRST}$\tablenotemark{a}} &
  \colhead{$f_{err, F}$} &
  \colhead{$f_{1.4GHz, NVSS}$} &
  \colhead{$f_{err, N}$} &
  \colhead{$f_{\nu, 2500}$} &
  \colhead{$f_{\nu, 4400}$} &
  \colhead{$f_{\nu, 5GHz}$} &
  \colhead{$R_{2500}$} &
  \colhead{$R_{4400}$} &
  \colhead{$M_{2500}$}
  }
\startdata
  J0011+1446 & 4.96 & 18.03 & 23.96 & 0.146 & 35.8 & 1.5 & 0.0491 & 0.0651 & 5.1933 & 105.8 & 79.7 & $-$28.65\\
  J0131$-$0321 & 5.18 & 18.09 & 32.83 & 0.123 & 31.4 & 1.0 & 0.0448 & 0.0594 & 6.9880 & 156.0 & 117.6 & $-$28.66\\
  J0741+2520 & 5.21 & 18.22 & 2.07 & 0.141 & $--$ & $--$ & 0.0396 & 0.0525 & 0.4395 & 11.1 & 8.4 & $-$28.54\\
  J0813+3508 & 4.92 & 19.17 & 20.04 & 0.156 & 35.6 & 1.1 & 0.0173 & 0.0229 & 4.3583 & 252.0 & 189.9 & $-$27.50\\
  J1146+4037 & 4.98 & 19.48 & 12.45 & 0.146 & 12.5 & 0.5 & 0.0129 & 0.0171 & 2.6940 & 209.3 & 157.8 & $-$27.21\\
  J1318+3418 & 4.82 & 19.09 & 3.73 & 0.148 & 3.5 & 0.4 & 0.0189 & 0.0251 & 0.8181 & 43.2 & 32.6 & $-$27.54\\
  J2329+3003 & 5.24 & 18.83 & $--$ & $--$ & 4.9 & 0.4 & 0.0224 & 0.0298 & 1.0380 & 46.2 & 34.9 & $-$27.94\\
  J2344+1653 & 5.00 & 18.54 & $--$ & $--$ & 15.3 & 0.6 & 0.0305 & 0.0404 & 3.3052 & 108.4 & 81.7 & $-$28.15
\enddata
\tablenotetext{a}{Flux density and flux density error are in units of mJy. }
\tablenotetext{b}{Data in this table are calculated based on cosmology $H_{0}$ = 70 km s$^{-1}$ Mpc$^{-1}$, $\Omega_{m}$ = 0.3, and $\Omega_{\Lambda}$ = 0.7}
\tablenotetext{c}{The two objects only detected by NVSS without FIRST detection are not covered by the FIRST footprint.}
\end{deluxetable*}

\begin{figure}
\includegraphics[width=0.5\textwidth]{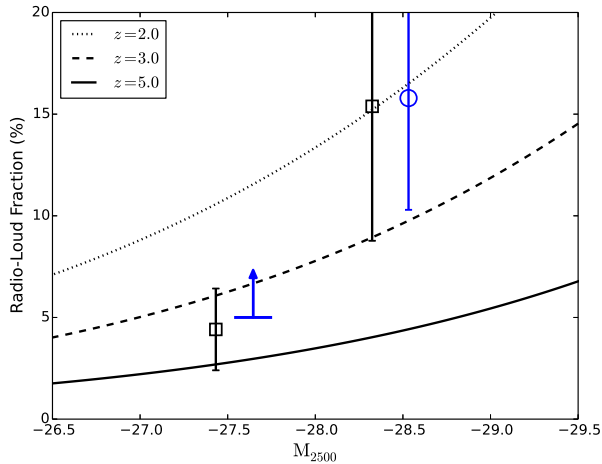}
\caption{The RLF in each magnitude bin compared with the evolution function from \cite{jiang07}. The dotted, dashed and solid lines show the predicted RLF at $z$ = 2, 3 and 5 from  \cite{jiang07}. The black squares represent the RLF in the FIRST area, and the blue circle and blue arrow denote the result from quasars within the NVSS area. In the faint magnitude bin, the RLF based on NVSS data is a lower limit. To show the points clearly, we shift the NVSS points by $-0.2$ mag. The uncertainties of RLF is estimated by assuming a poisson distribution of the number of quasars in each bin.} 
\label{fig10}
\end{figure}

\section{Conclusion}      
We establish a highly effective $z \sim$ 5 quasar selection method based on SDSS and ALLWISE optical/near-infrared colors. We relax the traditional $r-i/i-z$ color limit by including color cuts in the ALLWISE W1 and W2 data. We selected 110 quasar candidates that satisfied our selection criteria with good optical image quality and obtained spectroscopic observations for 99 candidates. 64 new quasars have been discovered in the redshift range $4.4 < z < 5.5$ and magnitude range $-29 < M_{1450} < $-$26.4$. We restrict our luminous quasar sample to $4.7 \le z < 5.4$ and $ M_{1450} \le $-$26.8$ for the QLF calculation. Combining all previously known quasars in this range, we construct the largest luminous quasar sample at $z \sim$ 5 and determine the QLF, covering a sky area of 14555 deg$^2$. Here we list our main conclusions. 
\begin{itemize}
             \item Within the redshift range $4.7 \le z < 5.4$ and magnitude range $ M_{1450} \le $-$26.8$, there are 45 newly identified quasars and 54 known quasars. Our new discovery successfully extends the population of luminous quasars at $z \sim$ 5, especially at $ M_{1450} \le $-$27.3$, where we discovered 27 new quasars and increased the number of known quasars by a factor of 1.5 in this luminosity range. Our final sample including 99 quasars is the largest sample of luminous $z \sim$ 5 quasars (Fig. 1).
             \item We derive the selection function of our color-color selection by using 311,000 simulated quasars in the redshift range $z$ = 4 to 6 and luminosity range $-$29.5 $< M_{1450} < $-25.5. The selection function shows that by relaxing the traditional $r-i$/$i-z$ color cut and adding the W1-W2 color, our color selection criteria extend the selection fuction to a higher redshift z$\sim$ 5.4 than previous work (Fig. 5).
             \item Using this sample, we calculate the binned QLF and fit the parametric QLF by using maximum likelihood fitting at $z = 5.05$ (Fig. 8 \& 9). For the parametric QLF, we fix the faint end slope $\alpha$ =-2.03, which is measured by using the S82 and DR7 quasar samples (M13), and find the best fit result of the bright end slope $\beta$ = $-$3.58$\pm$0.25 and  break magnitude of $M_{1450}^{*}$ = $-$26.99 $\pm$0.23. 
             \item We compare parameters of our best fit QLF with previous work at different redshifts and use all points to fit an LEDE model. Our result is consistent with the previous LEDE model but prefers a steeper slope of log$\Phi^{*}(z)$ evolution and a flatter slope of break magnitude evolution. The comparison for $\beta$ shows no clear evolution with redshift (Fig. 11). 
             \item We calculate the contribution of quasars to the ionizing background at $z \sim$ 5 based on our QLF. Integrating our best fit QLF, we find that quasars are able to provide $\sim$ 18\% $-$ 45\% of the required photons based on a clumping factor $C \sim$ 2 $-$ 5. 
             \item We use FIRST and NVSS data to calculate the radio loud fraction of our sample and give a lower limit for the RLF of $\sim$ 7.1\% which agrees with the result at $z \sim$ 6 of \cite{banados15}. In comparison with the predicted evolution function of the RLF with $M_{2500}$ and $z$ proposed by \cite{jiang07}, our result shows evolution with optical luminosity but no obvious evolution with redshift (Fig. 12).             
\end{itemize}

\acknowledgments
We thank the referee for providing constructive comments and suggestions. J. Yang, F. Wang and X.-B. Wu thank the supports by the NSFC grant No.11373008 and 11533001, the Strategic Priority Research Program ''The Emergence of Cosmological Structures'' of the Chinese Academy of Sciences, Grant No. XDB09000000, and the National Key Basic Research Program of China 2014CB845700. J. Yang, X. Fan and I. D. McGreer acknowledge the support from the US NSF grant AST 11-07682. Funding for the Lijiang 2.4m telescope is provided by Chinese Academy of Sciences and the People's Government of Yunnan Province. This research uses data obtained through the Telescope Access Program (TAP), which has been funded by the Strategic Priority Research Program "The Emergence of Cosmological Structures" (Grant No. XDB09000000), National Astronomical Observatories, Chinese Academy of Sciences, and the Special Fund for Astronomy from the Ministry of Finance in China. We acknowledge the use of the Lijiang 2.4 m telescope, the MMT 6.5 m telescope, the Bok telescope, ANU 2.3m telescope and Xinglong 2.16m telescope.  We acknowledge the support of the staff of the Xinglong 2.16m telescope. This work was partially supported by the Open Project Program of the Key Laboratory of Optical Astronomy, National Astronomical Observatories, Chinese Academy of Sciences. 

We acknowledge the use of SDSS photometric data. Funding for SDSS-III has been provided by the Alfred P. Sloan Foundation, the Participating Institutions, the National Science Foundation, and the U.S. Department of Energy Office of Science. The SDSS-III Web site is http://www.sdss3.org/. SDSS-III is managed by the Astrophysical Research Consortium for the Participating Institutions of the SDSS-III Collaboration including the University of Arizona, the Brazilian Participation Group, Brookhaven National Laboratory, University of Cambridge, Carnegie Mellon University, University of Florida, the French Participation Group, the German Participation Group, Harvard University, the Instituto de Astrofisica de Canarias, the Michigan State/Notre Dame/JINA Participation Group, Johns Hopkins University, Lawrence Berkeley National Laboratory, Max Planck Institute for Astrophysics, Max Planck Institute for Extraterrestrial Physics, New Mexico State University, New York University, Ohio State University, Pennsylvania State University, University of Portsmouth, Princeton University, the Spanish Participation Group, University of Tokyo, University of Utah, Vanderbilt University, University of Virginia, University of Washington, and Yale University. This publication makes use of data products from the Wide-field Infrared Survey Explorer, which is a joint project of the University of California, Los Angeles, and the Jet Propulsion Laboratory/California Institute of Technology, and NEOWISE, which is a project of the Jet Propulsion Laboratory/California Institute of Technology. WISE and NEOWISE are funded by the National Aeronautics and Space Administration.

{\it Facilities:} \facility{Sloan (SDSS)}, \facility{WISE}, \facility{2.4m/YNAO (YFOSC)}, \facility{MMT (Red Channel spectrograph)}, \facility{2.16m/NAOC (BFOSC)}, \facility{2.3m/ANU (WiFeS)}, \facility{Bok/Steward Observatory(B\&C)}.


\clearpage

\end{document}